\documentclass[conference]{IEEEtran}
\IEEEoverridecommandlockouts

\usepackage{cite}
\usepackage{amsmath,amssymb,amsfonts}
\usepackage{algorithmic}
\usepackage{graphicx}
\usepackage{textcomp}
\usepackage{listings}
\usepackage[bookmarks=false]{hyperref}
\usepackage{makecell}
\usepackage{enumitem}
\usepackage[super]{nth}
\usepackage{tikz}
\usetikzlibrary{shapes,positioning,calc}

\def\BibTeX{{\rm B\kern-.05em{\sc i\kern-.025em b}\kern-.08em
    T\kern-.1667em\lower.7ex\hbox{E}\kern-.125emX}}
\begin{document}

\newlist{legal}{enumerate}{10}
\setlist[legal]{label*=\arabic*.}

\lstset{
    language=C++,
    breaklines = true,
    upquote = true,
    columns = flexible,
    basicstyle = \ttfamily,
    frame = single,
    keepspaces = true,
}

\title{Dynamic Analysis of ARINC 653 RTOS \\with LLVM
}

\author{\IEEEauthorblockN{Vitaly Cheptsov}
\IEEEauthorblockA{\textit{Ivannikov Institute for System Programming}\\\textit{of the Russian Academy of Sciences;}\\\textit{Higher School of Economics} \\
Moscow, Russia \\
cheptsov@ispras.ru}
\and
\IEEEauthorblockN{Alexey Khoroshilov}
\IEEEauthorblockA{\textit{Ivannikov Institute for System Programming}\\\textit{of the Russian Academy of Sciences} \\
Moscow, Russia \\
khoroshilov@ispras.ru}
}

\maketitle

\begin{tikzpicture}[remember picture, overlay]
    \path node at ($(current page.south) + (0,0.65in)$) {
        \begin{minipage}{\textwidth} \footnotesize
            V. Cheptsov and A. Khoroshilov, ``Dynamic Analysis of ARINC 653 RTOS with LLVM,'' 2018 Ivannikov Ispras Open Conference (ISPRAS), 2018, pp. 9-15, DOI: \href{https://doi.org/10.1109/ISPRAS.2018.00009}{10.1109/ISPRAS.2018.00009}.

            \copyright~2018 IEEE. Personal use of this material is permitted. Permission
            from IEEE must be obtained for all other uses, in any current or future media,
            including reprinting/republishing this material for advertising or promotional
            purposes, creating new collective works, for resale or redistribution to
            servers or lists, or reuse of any copyrighted component of this work in other
            works.
        \end{minipage}
    };
\end{tikzpicture}
\begin{abstract}
Existing standards for airborne-embedded software systems impose a number of requirements applicable to the software development cycle of hard real-time operating systems found in modern aircraft. The measures taken are meant to reduce the risks of undesired consequences, but have strongly varying costs. Dynamic instrumentation and static analysis are common practices used to automatically find software defects, from strictly non-conforming code constructions to memory corruptions or invalid control flow. LLVM analyser and sanitizer infrastructure, while regularly applied to general-purpose software, originally was not thought to be introduced to heavily restricted environments. In this paper we discuss the specifics of airborne systems with regards to dynamic instrumentation and provide practical considerations to be taken into account for the effective use of general-purpose instrumentation tools. We bring a complete LLVM stack support to JetOS, a prospective onboard real-time operating system currently being developed at ISP RAS in collaboration with GosNIIAS. As an example, we port AddressSanitizer, MemorySanitizer, and UndefinedBehaviorSanitizer and provide the details against the caveats on all relevant sides: a sanitizer, a compiler, and an operating system. In addition we suggest uninvolved optimisations and enhancements to the runtimes to maximise the effects of the tools.
\end{abstract}

\begin{IEEEkeywords}
dynamic instrumentation, real-time operating systems, ARINC 653, IMA, LLVM
\end{IEEEkeywords}

\section{Introduction}
The problem of code correctness existed since before it was studied in computer science. As long as programming is done by humans, there will remain a high probability of mistakes in the written code, which may result in program malfunctioning at some point in the future. To circumvent biological imperfection, a number of methods reducing the risks of making mistakes during software development, decreasing the costs for their elimination, and, most importantly, detecting the mistakes made as early as possible, are currently being developed.

Depending on the software lifecycle and regulatory documents used, the set of measures taken may be comprised of: coding standards, test coverage, defensive programming, redundancy, semiformal and formal objectives to supporting documentation, threat model conformance testing, static analysis, dynamic instrumentation, fuzz-testing, regular inspection, code modelling and verification, etc. The minimal objectives for general-purpose software development may be defined by national and international standards for secure development lifecycle, like GOST R 56939—2016~\cite{gost_sec}.

Safety critical software is an entirely different domain. Creation of systems, failure of which costs human loss, injuries or disasters resulting in severe nature or property damage, may only be performed under a large amount of strict precautions, commonly referred to as objectives, defined in industrial standards. For aviation it is DO-178C~\cite{do_178c}. The resulting product has to be certified prior to being put into service, and history shows that the price of timely defect detection is quite different at the early stage versus the verification or production stage~\cite{jones}. To lower the costs, additional measures, sometimes not included in particular regulatory documents, are continuously sought for and efficiently applied whenever possible. This is especially important when certifying against multiple standards containing objectives which not necessarily conform to each other.

In this paper we study the specifics of modern airborne operating systems in regards to dynamic instrumentation for early error detection. We take a set of general-purpose error detection tools based on the LLVM sanitizer infrastructure, targeting dynamic software instrumentation and error detection, and bring them up in a certified airborne operating system JetOS~\cite{jetos}. These tools are commonly applied to general-purpose software development projects but were not intended to be used in restricted environments. We describe which tools may be applicable for testing ARINC 653 applications and the operating system itself, which objectives and issues could be focused on, and the specifics of implementing sanitizer runtime support in a hard real-time operating system. We showcase possible issues to be found and suggest actions to increase the efficiency of test results in tandem with the other tools.

\section{State of the Art} \label{subsec:stateofart}

Dynamic instrumentation is a way to measure program behaviour by certain parameters allowing to gather information about program performance, contract violation, coverage, and debugging at runtime. Different kinds of instrumentation may require code alteration on the source level or the compiler level. In this paper we refer to dynamic instrumentation in the sense of compiler passes applying modifications targeting programming error detection.

While the benefits of dynamic instrumentation may be vast, there are not so many publicly available tools that could be integrated at a reasonable cost. Proprietary software, such as WinDbg, Intel Inspector, UNICOM PurifyPlus, Insure++, and several others, usually target specific platforms or userspace instrumentation, and even if free for use come with no source availability. As a result, it is either impossible or very costly to adopt it to embedded projects, putting aside the safety-critical systems segment. The few existing opensource tools have the exact same problems except in-house resources are to be spent for the bring up.

Valgrind is one of the oldest and most powerful tool suites currently developed. It supports a large number of CPUs and was proven to be quite effective on general-purpose operating systems, like Linux targets, macOS, or Android. However, using Valgrind on non-Linux and even non-POSIX targets may be quite challenging. One of the ways to use Valgrind could be porting the code to a simulated Linux-based environment, but this results in additional costs of supporting multiple systems as well as the inability to detect bugs in the original code, which makes the task effectively unrewarding. In addition, Valgrind seriously affects the performance of the instrumented code~\cite{valgrind}. DynamoRIO based Dr. Memory and the supplemental tools are quite promising, but once again the current development course is focused on expanding the functionality and resolving the issues in general-purpose operating systems. PowerPC or MIPS-based CPUs are not yet even considered.

From this perspective the LLVM project~\cite{llvm} favourably differs from the others by consisting of reusable modules handling compiling, building, and instrumenting stages altogether. Just like GNU Compiler Collection (GCC), LLVM subprojects contain all the necessary components to create a fully-fledged toolchain for a target Board Support Package (BSP). It has an optimizer with code generation support with broad target architecture and CPU compatibility, a C/C++/Objective-C code compiler, a set of libraries for lower-level code generation support, C++ standard library, as well as its own linker and debugger.

The LLVM project additionally contains a large set of dynamic instrumentation tools. Most of them were originally developed by Google Sanitizers~\cite{google_san} project, and while they require runtime support from the target platform, this runtime can be written in C or C++, and all the checks necessary are automatically inserted by the compiler, so they do not require additional assembly usage. Among the currently supported tools are the following:

\begin{itemize}
\item AddressSanitizer – fast memory usage error detection tool. Used for buffer overflow errors, memory reuse after free, and so on;
\item MemorySanitizer – uninitialised memory usage error detection tool;
\item UndefinedBehaviorSanitizer – undefined behaviour detection tool. Used for locating the use of unaligned pointers, integer overflow, incorrect execution of floating-point calculations, incorrect boolean type assignment, etc;
\item ThreadSanitizer – data race condition detection tool. Used for finding errors in multithreaded programs;
\item Control Flow Integrity – tool implementing a number of control flow integrity checks including the detection of incorrect class and function casts and invalid virtual table usage;
\item LeakSanitizer – memory leak detection tool;
\item EfficiencySanitizer – tool for detecting suboptimal language constructions and APIs;
\item SafeStack – tool intended to protect software from stack buffer overflow with no measurable performance costs;
\item libFuzzer – fuzz testing tool. 
\end{itemize}

These tools are being continuously ported to GCC as well, however, the GCC support list is incomplete and suffers from backport delays. On the good side, LLVM and Clang try to support a reasonable amount of GCC-supported platforms and extensions, which allows a drop-in replacement in the majority of the BSPs. 

\section{JetOS Operating System}

JetOS is a cross-platform hard real-time operating system designed to control airborne equipment in modern civil aviation. One of the fundamental requirements in its implementation is accordance with ARINC 653~\cite{arinc653} parts 1 and 2. ARINC 653 (Avionics Application Standard Software Interface) describes the implementation of APEX (APplication/EXecutive), an API that regulates space and time partitioning in safety-critical real-time operating systems for avionics. This standard allows the hosting of multiple applications of different software levels on the same hardware in the context of an Integrated Modular Avionics (IMA)~\cite{ima_ARI91} architecture. In turn, IMA is an airborne system architecture used in modern aircrafts (Boeing 787, Airbus A350, etc.) replacing the outdated Federated Architecture (FA). It features a common network of software components portable among standard hardware modules, which allows to significantly reduce hardware resource consumption.  

As per ARINC 653, airborne OS subsystems are divided into isolated dedicated sections called partitions, which run cyclically following a predefined schedule. Each partition houses one or more application process, varying in durability, period, priority, and current state. Inter partition communication is executed through ARINC 653 ports in the form of a FIFO queue (queueing messages) or only reading the latest arrived message (sampling messages). If partitions are placed at different processor modules the communication happens over buses or switched networks like Avionics Full-Duplex Switched Ethernet (AFDX)~\cite{afdx}.

DO-178C standard, which regulates software development in civil aviation domain, defines a number of requirements. It covers all the processes including planning, development, verification, configuration management, quality assurance, and certification. The set of requirements depends on the software severity level that ranges from level~A (failure results in catastrophic consequences and loss of an aircraft) to level~D (failure results in minor discomfort for passengers and crew)~\cite{do178c_compl}. According to the DO-178C, development of any critical software starts from system level requirements. Afterwards high-level software requirements are described and software architecture defines decomposition to software subcomponents, which get low-level requirements. Later onwards the source code is developed, requirements-based unit and integration tests are written and executed, structural code coverage is measured and analysed. Moreover, at each development step Configuration Control requires that any modification in the first class development artifacts (like requirements and code) has to be justified and logged, which results in a repetition of all the verification activities. That makes the development process quite costly, especially if many repeated modifications are required.

Due to the aforementioned specialties, it is advisable to approach the development of operating systems of this class in two stages. At the first stage, a working prototype is created and used to form and implement the required functionality. The process permits testing multiple solutions to particular problems and helps to decide on the component development within the certification package. At the second stage, high-level requirements, architecture and design documentation, low-level requirements, and source code of the production version are developed in complete compliance with the certification process on top of the prototyping experience. This method allows to reduce the number of code modifications, increases the development efficiency, and lowers the costs.

At the time dynamic instrumentation was considered, JetOS was in prototype stage built on top of POK~\cite{pok_os} operating system. The size of the codebase was around 40000 LOC written in a subset of freestanding C99 with around 15000 LOC comprising the kernel code. The prototype was meant to be portable, and supported multiple architectures including:
\begin{itemize}
\item PowerPC 32-bit (e500mc, e500v2);
\item ARMv7 (Cortex-A7, Cortex-A9);
\item MIPS (MIPS32r1);
\item X86 (i486-compatible).
\end{itemize}

The buildsystem was already adopted to use a GCC-based toolchain, and suggested the possible use of certified backends in the future. All kinds of user-configurable compiler optimisations were prohibited, as well as compiler extensions for maximum compatibility with non-GCC compilers, like formally verified COMPCERT~\cite{compcert}. There are two supported configurations of the codebase:
\begin{itemize}
\item Onboard — the code intended to be executed on board of an aircraft, subject to formal inspection and later certification, part of the certification package;
\item Ground — the code not executed on board of an aircraft, used for the development and debugging of Onboard code, prototyping, training the staff, instrumentation, telemetry, and other needs.
\end{itemize}

The userspace environment is comprised of system code, meant to provide the necessary APIs and communicate with the kernel, and application software provided by various partners participating in the project. The following APIs were implemented or were in progress of being supported for application software:
\begin{itemize}
\item ARINC 653;
\item C Standard Library (limited subset);
\item OpenGL SC 2.0;
\item Embedded C++;
\item Ada Standard Library (limited subset).
\end{itemize}

\section{Dynamic Instrumentation Value}

Among the implemented measures of the secure development lifecycle JetOS already has APEX conformance tests, structural code coverage collection, unit tests for internal functions, static analysis, and a number of side tests for the general-purpose library interfaces like C standard library. Adding the support of dynamic instrumentation was a logical step forward, but the choice of LLVM was not unbiased. JetOS code is written in a portable manner without relying on compiler specifics, needs to target a wide range of architectures and processors, was buildable with GCC, and relied on LLVM static analysis tools, and most importantly was at considerably early development stage.

There are two main kinds of value provided by error detection techniques based on dynamic instrumentation:

\begin{itemize}
\item Reducing the debugging efforts by exposition of the place where the error happens in comparison to the place where a consequent fault became visible.
\item Detection of latent errors that do not manifest themselves in the tested conditions and configurations, but can be revealed in slightly altered environment or different preconditions.
\end{itemize}

For the first case the value comes from the ability of the tools to propagate the information of control and data flows to later stages of program execution. For instance, uninitialised pointer cannot be strictly considered as an error until it actually is dereferenced, but even then the results of this operation are not guaranteed to be instantly visible for the programmer. A tool that is able to track and find the origin of a sequence of unobvious memory corruptions or some other events may seriously increase the debugging speed. 

To illustrate the benefits of the second kind, let us give a generic example based on the code cited in \hyperref[asan1]{Listing 1}. On most platforms this code will not cause an abnormal termination or signal any problems, despite the presence of an undefined behaviour due to buffer overflow. If the buffer size was odd, some implementations would not even require \verb!buffer_prev! or \verb!buffer_next!, because the compiler could allocate the needed space implicitly while complying with the alignment requirements. Dynamic instrumentation is capable of detecting these and other `quiet' issues, which under certain conditions may lead to severe consequences. That said, dynamic instrumentation does not provide a complete guarantee of code correctness, and in most cases it only helps to detect problems with memory usage, synchronisation primitives, performing calculations etc., which could otherwise lead to a number of software errors if violated. Dynamic instrumentation in combination with other means has helped to uncover a substantial amount of previously undetected errors in existing open source projects~\cite{asan_chromium}~\cite{asan_linux} and has helped to prevent the occurrence of new ones, thus proving itself efficient. 

\begin{lstlisting}[caption=Generic buffer overflow example, frame=single, captionpos=b, label=asan1, float]
void test(void) {
    char buffer_prev[16];
    char buffer[16];
    char buffer_next[16];
    size_t sz = sizeof(buffer);
    printf("Local buffers %p %p %p\n",
      buffer_prev, buffer, buffer_next);
    *((volatile char *)buffer - 1) = 1;
    *((volatile char *)buffer + sz) = 1;
}
\end{lstlisting}

There are certain types of errors irrelevant to JetOS code. For example, memory cannot be reused after deallocation as there is no way to deallocate memory in JetOS except for areas solely managed by applications. 
However, the other types of errors may still happen.

Dynamic instrumentation could be useful to detect errors in JetOS during the both phases of its development.

\begin{enumerate}
\item The first stage of prototyping and porting components;
\item The later stage of production code development.
\end{enumerate}

The first stage has a lot in common with general purpose software. But as far as productivity of error detection based on dynamic instrumentation highly depends on thoroughness of applied tests, the more tests became available during the second stage the more possible situations are checked by instrumented code.

\section{Bringing Dynamic Instrumentation Support}

There are several dynamic instrumentation tools of the LLVM project and each of them has distinct purpose and specific requirements. Some of the tools are not so useful for JetOS:

\begin{itemize}
\item LeakSanitizer is architecturally inapplicable for ARINC 653 code, as the whole memory configuration is static, and each component owns a specific chunk of memory during the partition lifetime. It is possible to write dynamic memory management code for execution on the ground, but in this case there is no reason to use LeakSanitizer to detect memory leaks, as the allocator can have the necessary checks inside.
\item SafeStack targets stack corruption detection in production code. It implements a small subset of checks available in AddressSanitizer but in a very efficient way. SafeStack is useful for general purpose software where typical test coverage does not ensure that all possible problems were detected during testing even with AddressSanitizer enabled. While there is no absolute assurance of error detection during testing in avionics as well, but more thorough test coverage reduces the value of the SafeStack especially taking into account additional CPU and verification overhead coming with it.
\item Control Flow Integrity is not beneficial, since it provides C++ oriented checks, and C++ language is not common in onboard RTOS applications.
\end{itemize}

Other tools do not contradict the environment prerequisites, but have varying benefits. In JetOS our primary target was to provide the necessary tools to help the developers of application software. Taking the available resources we decided to focus on AddressSanitizer, MemorySanitizer, and UndefinedBehaviorSanitizer as our primary goals, and libFuzzer and ThreadSanitizer as our secondary goals. Once implementing them in userspace, we ported them to the kernel for the additional use. The reasons behind the decision were that the first three tools complement each other, and are often used together accompanied by the fuzzing, which could be initially substituted by a large set of tests with decent coverage.

In general case adopting LLVM dynamic instrumentation support consists of two parts:

\begin{itemize}
\item Compiler side support with instrumented code generation;
\item Target platform side support providing the necessary library API and other interfaces necessary for the tool functioning.
\end{itemize}

In LLVM each sanitizer has its own scope of supported architectures and target platforms. Even though, provided a functional runtime library implementation, there may be no technical limitations to run MemorySanitizer on PowerPC bare metal, Clang frontend may not allow emitting code with sanitizer passes. For this reason, it should be decided whether a separate target configurable to the needs of a project is to be created, or a general-purpose target (e.g. Linux) could be used. For JetOS it was possible to rely on stock LLVM and Clang by maintaining GNU ABI compatibility and use Linux target with upstreamed patches unlocking sanitizer support on all architectures assuming the runtime library is provided.

There are several approaches to implement target platform support as well. There is LLVM compiler-rt project, which offers a reference implementation for each tool written in C++. It supports a large number of general-purpose operating systems, notably Linux, macOS, BSD variants, Fuchsia, and Windows. While these implementations do not require C++ standard library availability, they still need some C standard library functions and headers, language features like thread-local variables, and a platform specific part providing stack unwinding, memory manipulation, synchronisation primitives, and several others. As an advantage, the reference implementation is always up-to-date, and requires few changes to maintain once it works. An alternative approach is to take a third-party runtime or implement it from scratch.

Since our main target was instrumenting userspace code, and optional C++ support was already considered for some of the tools, we opted to reuse the reference implementation where possible. For the tools used in both user and kernel space we opted with in-house implementations. The instrumentation runtime dependencies were minimised by integrating the existing services of JetOS debugging facilities. Unlike generic software, error discovery rate in life critical software is very low. Dynamic instrumentation serves as a proof of no conditions for certain error classes under certain circumstances. For this reason, providing extensive debugging facilities for report generation may not be beneficial. Adopting the runtime to be connected to the debugger scripts may be sufficient, since it already provides stack unwinding, variable printing, and source mapping to help locate the error.

Besides automatically generated checks, different tools provide an API, which may be used to implement sanity checks otherwise impossible to do directly in the codebase. Since it is not an option to fill safety-critical software sources with numerous conditionally enabled functional macros invoking sanitizer checks, we had to worked out several ways of adding the instrumentation intrinsics:

\begin{itemize}
\item Templated code generation — for the generated sources we have the necessary intrinsics right within the template, enabled when building with sanitizer support.
\item Function wrapping — select functions are provided with external wrapping, and their calls are redirected to a generated proxy, which itself calls the original function. This is problematic if it requires to maintain two separate interfaces and be aware of interface internals, but works fairly well for standardised functions.
\item Per-function preprocessing — based on a set of rules selected source files may be patched with the necessary intrinsics at designated places prior to compiling with sanitizer support. At this step we plan to automate the process of generating rules based on function specifications we have written.
\end{itemize}

\subsection{Implementing AddressSanitizer}

The general idea behind AddressSanitizer~\cite{asan} is to map conventional memory to a special area called shadow memory, which contains memory usage status rather than specific values. This tool detects errors like out of bounds access, unallocated memory access, out of the scope memory usage, stack corruption, as well as other related memory problems. Anytime conventional memory is addressed, the compiler generates a check of the corresponding area of shadow memory, and in the event of attempted access to invalid memory, signaled by poisoned shadow memory zones, a fatal exception is raised. Besides the practical benefit of the tool, coverage of several DO-178C Requirements-Based Testing Methods (6.4.3~\cite{do_178c}) objectives is also provided. The testing uncovers stack corruption, software partition violations, data corruption, and invalid parameter passing.

The main problem with using dynamic instrumentation and AddressSanitizer in particular for analysis of real-time software is timing overhead expenses. The aforementioned tools typically cause a double slowdown of code execution, which is negligible in general-purpose operating systems but in the case of RTOS may cause the code to go off-schedule and render software testing impossible. Potential solutions include local timeout alterations or a global timer delay in the entire operating system. In JetOS we had to apply both. In most cases it was sufficient to divide the timer output by some constant factor, pretending that it goes e.g. twice slower, but in some places the timeouts still had to be altered locally. The reason for global timer slowdown not working well enough is because the estimated slowdown caused by instrumentation and mentioned in the docs is usually average and approximated, while in reality it strongly varies on the code path. All in all, when dynamically instrumenting RTOS the reliability of the timing tests has to be given up leaving just the failure reproducibility as an important factor.

Another traditional problem for operating systems of this class is memory usage, which is allocated statically before the start of each partition and cannot be increased after that. To resolve this, JetOS uses a mechanism of preliminary memory area designation for each process, based on which the needed shadow memory size is calculated and provided to AddressSanitizer. Shadow memory granularity is one of the practical sides of AddressSanitizer. It allows to reduce the shadow memory size by mapping multiple origin bytes to one shadow byte with little to no report quality loss. The default granularity for AddressSanitizer is 8:1, which means 8 origin bytes are mapped to 8 shadow bits. Since the whole software package was under our control, we decided to reconsider AddressSanitizer memory checking from just heap, registered data variables, and stack to overall memory as a whole, blacklisting everything but what we are aware of like MMIO areas, configuration memory blocks, shared memory, etc. In our case sparse shadow memory or algorithm changes were not required, as we had enough hardware resources, but the caveats of working on minimalistic environment are mentioned in other works such as Address Sanitizer on Myraid~\cite{miraid}.

\subsection{Implementing MemorySanitizer}

MemorySanitizer~\cite{msan} is similar to AddressSanitizer but it targets uninitialised memory usage of heap and stack variables. Unlike AddressSanitizer, MemorySanitizer only supports 1:1 memory granularity, which means its shadow size must be equal to the used memory size. DO-178C Requirements-Based Testing Methods (6.4.3~\cite{do_178c}) specifies a number of objectives which are also applicable to MemorySanitizer related to variable initialisation correctness and valid parameter passing.

MemorySanitizer may not be directly compatible with a certified onboard OS. In safety-critical software it is not uncommon to circumvent variable initialisation issues by always initialising every declared variable with a special reserved initialisation value, when the actual value cannot be determined beforehand. It is not allowed to intentionally compare the variable value against this reserved value, assume it to be a valid state of a variable, or pass it to external functions, but it can be used as a last resort of control to make the behaviour deterministic and protect against initialisation errors and control flow attacks. Since the variable still has to be initialised by a correct value regardless of the reserved value presence, we opted to preprocess the sources with an external tool to remove the reserved initialisation value assignments. Another way avoiding source modifications could have been to modify the tool itself to account for the default value, but this would have caused various compatibility issues with uncertified software prototypes, which may known about the reserved initialisation value, and will lead to the loss of bit-precise error detection.

Memory returned from an unknown source, such as kernel memory through a syscall should be annotated to unpoison (mark as valid) output parameters. The example provided in \hyperref[msan1]{Listing 2} shows an annotated syscall template for later C conversion with the checks of input parameter initialisation and unpoisoning of the output parameters in case of operation success.

\begin{lstlisting}[caption=Annotated syscall example, frame=single, captionpos=b, label=msan1, float]
//!USER_NAME: jet_thread_status
//!PRE: msan_check (
    &thread_id, sizeof (thread_id));
//!POST: msan_unpoison (
    name, sizeof (max_name_t));
//!POST: msan_unpoison (
    entry, sizeof (*entry));
//!POST: msan_unpoison (
    status, sizeof (*status));
syscall_declare (
   jet_syscall_thread_status_t,
   jet_thread_get_status,
   jet_thread_id_t, thread_id,
   max_name_t, name,
   void**, entry,
   jet_thread_status_t*, status)
\end{lstlisting}

The same is being done in JetOS, not just to syscalls, but other functions based on the formally defined requirements and serving for contract enforcement as mentioned previously. One of the issues we had to workaround when using MemorySanitizer was providing annotations to structure types with padding. Structure padding may not be initialised and can cause false positives. Since the compiler is aware of the underlying type, we believe it should be possible to provide an API that will not require writing assertions field by field.

During the process of adding the checks we were able to identify select issues in the examples and test suite. For instance, one of the checks exposed a change in the \nth{4} revision of the \nth{1} part of ARINC 653 specification that was not reflected in the prototype. Earlier revisions of the standard permitted \verb!GET_MY_ID! to return \verb!INVALID_MODE! for the main process, since it had no ID assigned, while the latest revision defined \verb!MAIN_PROCESS_ID! for this purpose. While the complaint is not fair, as the test suite and prototype standard revisions were not fully coordinated, it makes us believe that sanitizer-aided assertions will serve as a helpful instrument to application software developers on both software and requirement levels.

\subsection{Implementing UndefinedBehaviorSanitizer}

UndefinedBehaviorSanitizer~\cite{ubsan} is an undefined behaviour detection tool, which helps to reveal errors such as misaligned or null pointer access, integer overflows, array bound violations when statically determinable, floating point issues such as division by zero and cast overflows, \verb!_Bool! or \verb!enum! range violations, integer truncation, etc.

While the simplest to support, UndefinedBehaviorSanitizer is one of the most powerful checkers of the toolset. It does not require special features like shadow memory, and currently three separate publicly available implementations exist: reference, Linux Kernel, NetBSD. In JetOS we relied on the reference implementation in userspace, and on NetBSD implementation in the kernel. It is discussable how dangerous the errors found can actually be, but for instance, in the prototype we discovered a long-standing bug in the assembly code of one of the BSPs, which did not evince itself in any way in the past but caused undefined behaviour in C code. Nullability checks guided by the \verb!_Nonnull! keyword can also serve as an enforcement of function requirements similarly to AddressSanitizer and MemorySanitizer range checks.

\section{Conclusion}


Based on our experience with JetOS prototype, we affirm that even though the instrumentation error detection rate may be low, the errors detected could be remarkably hard to discover with the other measures.

By the end of this work, we were able to prove the usefulness of general-purpose dynamic instrumentation tools during the development of ARINC 653 compatible safety-critical real-time operating system prototype. The benefits of LLVM sanitizers in prototyped and ported components are similar to those of non-specific software including early stage error detection as well as time and cost reduction. Furthermore, with almost no additional work, these tools can also be used for objective enforcement and contract checking of the certified code. Dynamic instrumentation could be viewed as automatic oracles that detect certain classes of errors during test execution and a testing method providing the guarantees of certain class error absence on the path, which in conjunction with high coverage, serves as a solid proof that all the other measures taken function correctly. 

In the continuation of our work we foresee several directions to consider. With AddressSanitizer, MemorySanitizer, and UndefinedBehaviorSanitizer already proven to be effective over the existing featureset, the works to expand their functionality involving upstream code modifications and downstream annotations can now be considered. We also plan to find the use of more tools. For instance, fuzzing or symbolic-execution based fuzzing could quickly expand the coverage area when the test suite is not yet complete, which is generally useful for the prototyped components or testing kernelspace isolation. Unlike sanitizers, which cause undesired but measurable slowdown, fuzzing is an entirely different domain and may be a lot more complex to support in a real-time operating system. However, we suppose that even with a well-written test suite fuzzing backed with sanitizer instrumentation may uncover new code paths or different behaviour of the old paths, and further research of the topic is worth the effort. The primary fuzzing targets could be AFDX network, device communication, and system calls. Regarding the AFDX network used for inter-partition communication, at present the target partition has to determine the validity of transmitted data based on the message contents, however, the source partition already has parts of this information from the sanitizer runtime. It verifies the properties like initialisation of the transmitted data and enforces them prior to sending. Whether or not the possibility to transmit supplemental information about the contents of the message would be beneficial is also questionable.

Overhead induced by dynamic instrumentation leads to issues with real-time scheduling. We overcome it by introducing timer slowdown option, but the problem of estimation of timing overhead and making it deterministic is an open question for further research.


\section*{Acknowledgements}

This study is supported by RFBR grant \#16-01-00356.

\IEEEtriggeratref{4}

\end{document}